# SUPERSHARP

**Segmented Unfolding Primary for Exoplanet  Research via Spectroscopic High Angular Resolution Photography**

Lead proposer: Ian Parry
University of Cambridge, UK

Core team members: D. Queloz, G. Kennedy, N. Madhusudhan, A. Triaud, N. Walton, R. Vasudevan, P. Zulawski (Cambridge, UK), K. Heng, W. Benz, C. Mordasini, N. Thomas,  D. Piazza (Bern, CH), S. Udry, (Geneva, CH), S. Quanz, (ETH, CH), D. Mouillet,  J.-L. Beuzit (Grenoble, F), I. Snellen, M. Kenworthy (Leiden, NL), D. Pollaco (Warwick, UK), S. Hinkley (Exeter, UK), B. Biller (Edinburgh, UK), S. Rugheimer (St Andrews, UK).

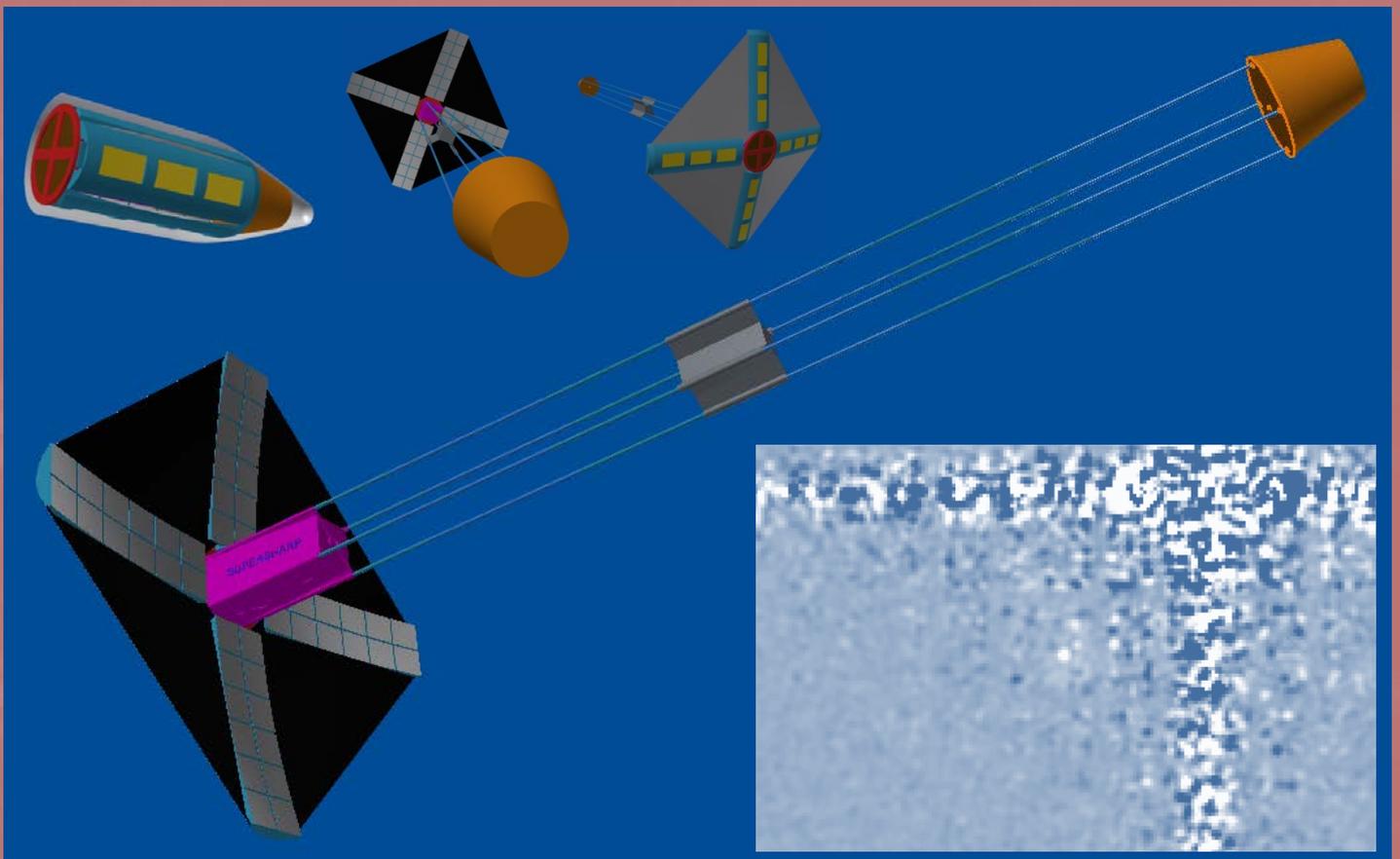

# 1) Executive Summary

How common is life in the Universe? This is one of today's biggest scientific questions but we do not have an answer: all we can say is that there is at least one life-bearing planet in the entire universe. ESA, of course, has long recognised that this is a vitally important question and it is a cornerstone of ESA's Cosmic vision program (COSMIC VISION 2015-2025: PLANETS AND LIFE. Theme 1.2 "from exo-planets to biomarkers". Goal: Search for planets around stars other than the Sun, looking for biomarkers in their atmospheres, and image them). In this proposal we present this as the "new science" that needs to be done and we describe SUPERSHARP which is our concept for how it can be done.

Specifically, we are proposing to search for biosignatures in the spectra of reflected light from ~100 Earth-sized planets that are already known to be orbiting in their habitable zones (HZ). For a sample of G and K type hosts, most of these planets will be between 25 and 50 milli-arcsec (mas) from their host star and $10^{-9}$ to $10^{-10}$ times fainter. To separate the planet's image from that of its host star at the wavelength (763nm) of the $O_2$ biosignature we need a telescope with an aperture of ~16m. Furthermore, the intensity of the light from the host star at the position in the image of the exoplanet must be suppressed otherwise the exoplanet will be lost in the glare.

This presents huge technical challenges. The Earth's atmosphere is turbulent which makes it impossible to achieve the required contrast from the ground at 763nm. The telescope therefore needs to be in space and to fit the telescope in the rocket fairing it must be a factor of 4× or more smaller when folded than when operational. To obtain spectroscopy of the planet's biosignature at 763nm we need to use an integral field spectrometer (IFS) with a field of view (FOV) of 1000 × 1000 milli-arcsec (mas) and a spectral resolution of R~100. This is a device that simultaneously takes many pictures of the exoplanet each at a slightly different wavelength which are then recorded as a data cube with two spatial dimensions and one wavelength dimension. In every data cube wavelength slice, the background light from the host star at the location of the planet image must be minimised. This is achieved via a coronagraph which blocks the light from the host star and active/adaptive optics techniques which continuously maintain very high accuracy optical alignment to make the images as sharp as possible. These are the technical challenges to be addressed in a design study.

We believe the cost of such a mission should not be estimated from the cost of other missions by scaling the telescope aperture and that **it can be affordable if its design is specifically aimed only at the exoplanet research goals.** Cost control will be a key part of future design studies. We are ultimately aiming for the L4 slot with a launch around 2039 and a total cost of €1B - €1.5B. We propose a staged approach to get to L4 adoption. First we need design studies, initially conceptual ones followed by actual lab prototypes. Then we need and an in-orbit small scale technology demonstrator.

Although our motivation to develop the technology for making large affordable space telescopes comes from a desire to search for life, the technology will have a great impact generally in astronomy and Earth observations (EO) with the latter also having great potential for commercial exploitation.

# 2) Main Science case – robustly searching for biosignatures

## Introduction

The key "new science" that needs to be enabled is the ability to detect biosignatures in the spectra of reflected light from a sample of Earth-sized exoplanets orbiting in their habitable zones. A sample size of ~100 is needed to begin to place robust constraints on the frequency of life in the universe (a quantity which is essentially unconstrained at the moment). To do this we need significant technology  development because it demands imaging-spectroscopy with very **high spatial resolution** and very **high contrast** to separate the light of the very faint planet from that of its very bright parent star which is only a few milli-arcsec away from it on the sky.

From our experience with ground-based instruments such as SPHERE and our understanding of the occurrence rates of exoplanets we essentially now know how to do this. The main limitation for a ground-based system such as SPHERE on the 8m VLT is the turbulence of the Earth's atmosphere which severely limits the contrast at the wavelength of the main biosignature, $O_2$ at 763nm. Even for the 39m E-ELT the atmospheric turbulence still prevents the required resolution and contrast from being achieved. Clearly, our successful ground based techniques need to be applied to a space mission to avoid the turbulence of the Earth's atmosphere and so we have developed a mission concept called SUPERSHARP (Segmented Unfolding Primary for Exoplanet  Research via Spectroscopic High Angular Resolution Photography) which is described below.

However, some of the key technologies needed for SUPERSHARP are at a low TRL (technology readiness level) for space applications and they need to be taken up to TRL 6 before a credible mission proposal can be made. In particular there are two technologies which need to be addressed in an imminent design study:

1) **Large foldable and deployable telescopes.** The high angular resolution requires a telescope which is much bigger than the rocket fairings of today's largest launch vehicles such as Ariane 6. The telescope therefore must be folded up for the launch and then deploy into a much larger structure once in space. This requires a deployable telescope structure and a segmented primary mirror. The bigger the expansion factor the better the science that is enabled.
2) **Active control of the optics and the minimisation of wave front error (WFE).** Once deployed, the optics (telescope + coronagraph) have to be aligned with great precision and this alignment must be continuously maintained. To achieve a low background from the parent star at the position of the planet the WFE must be very small. Furthermore, it must be possible to subtract off this residual background without systematic errors.

The deployable telescope structure technology specifically needed for SUPERSHARP is probably only at TRL 1 or 2 at the moment. However similar concepts have been studied and flown. JWST has an unfolding primary and an unfolding telescope structure but it has a much smaller expansion factor than what is needed for SUPERSHARP. JWST also has a segmented primary mirror.

An extending structure was studied by ESA for IXO (which later became ATHENA - CDF Study Report IXO Telescope - ESA Concurrent Design Facility - CDF-86(A) April 2009). Deployable booms have been used for many space application including large antennas, solar panels, sun shades and solar sails.

The active/adaptive control of optics is probably currently at TRL 4 or 5. Deformable mirrors (DM) have not yet been flown in space mainly because there's no atmospheric turbulence to correct. However, DMs will be used on WFIRST-AFTA to achieve excellent contrast performance and so will get to TRL 9 before 2030. Similarly, a DM will be extremely useful for SUPERSHARP to correct residual errors from the primary mirror and deal with any fast time frequency wavefront errors.

There is also a big concern about the cost. How can a huge telescope (>8m aperture) not cost much more than say Herschel? It is very tempting to think that the cost will be well over an L-class budget on the basis of some simple telescope size scaling law. The design studies must therefore show convincingly that such an approach to cost estimation is completely inappropriate. For example, missions to the outer solar system need much bigger solar panels but clearly it is wrong to estimate the cost of such a mission based on the size of the solar panels. Many of the mission costs for SUPERSHARP will be similar to those of previous M-class missions (for example the cost of the power system, the communication system or the operations phase). Some costs will be less than for previous missions (e.g. the launch cost for an Ariane 6 is less than for an Ariane 5). The key question therefore is how much extra cost arises through the two new technologies (i.e. deployment and WFE control) and this must be thoroughly addressed in the design studies.

This proposal specifies the science goals, identifies the technical requirements and describes a possible way forward to address both the TRL and cost issues. Our ultimate aim is to achieve the biosignature science goals within the constraints of the L4 mission.

## Direct imaging/spectroscopy – present and future

Figure 1 shows the limiting contrast as a function of angular separation for various telescopes and high spatial resolution imagers. This figure shows that no existing or under-construction facilities can achieve the required contrast and angular resolution combination (the grey shaded area) to do a HZ Earth-sized biosignature search.

Current state-of-the-art ground based systems (SPHERE, GPI, P1640, SCExAO) are limited by the Earth's atmosphere - their contrast is limited by the residual WFE due to atmospheric turbulence. The 5-σ contrast limit plotted in figure 1 is for 1 hour exposures. If the systematic errors that plague speckle subtraction can be reduced then better contrasts may be possible via longer exposures.

Even though HST has coronagraphic modes it still has relatively poor contrast performance compared to state-of-the-art ground-based coronagraphs. However, its performance extends to good angular resolution because it can operate at short wavelengths.

WFIRST-AFTA has the same primary mirror size as HST but will have much better contrast performance because it will have a state-of-the-art coronagraph. However, its primary mirror is too small (i.e. its IWA is too big) to resolve more than a few HZ Earth-size planets.

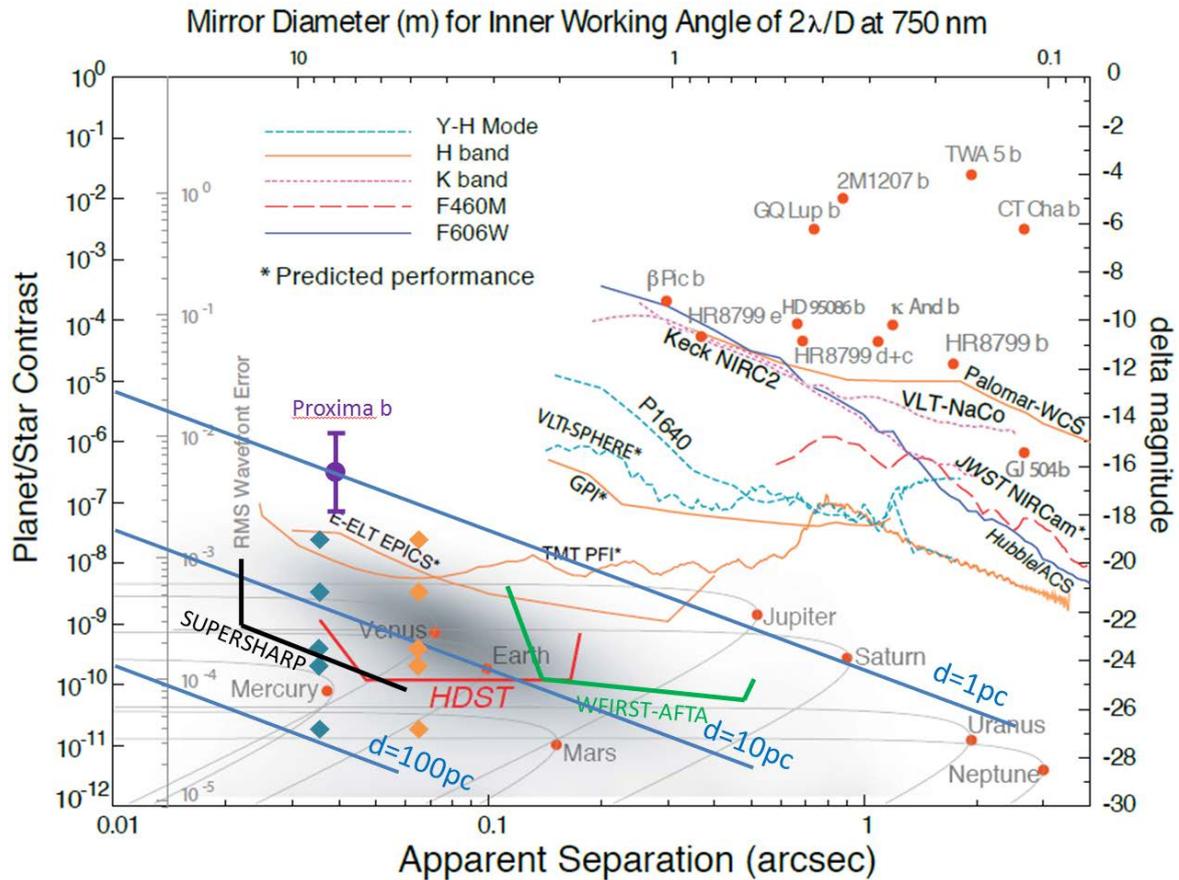

Figure 1: Contrast v apparent angular separation. Shown are the 5-σ contrast limits after post-processing one hour's worth of data for various coronagraph instruments. The WFIRST-AFTA lines in green are for 560nm. The recently announced Proxima b exoplanet is also shown. The 3 straight blue lines are for Earth-size planets at 3 different distances (this is the purple Earth line plotted in figure 7). The points in table 1 are shown as orange diamonds and those for table 2 as green diamonds. On the top right of the figure are plotted the K-band contrasts of some of the giant self-luminous exoplanets imaged to date. Current ground-based direct imagers only operate in the top right-hand corner. In the lower part of the figure are plotted our Solar System planets as they would appear in reflected light around a star at a distance of 10 pc. The region above the solid red line would be probed by HDST. Preliminary estimates for the limits of SUPERSHARP are plotted in black. The grey region in the lower left of the figure shows the predicted locus of terrestrial habitable zone planets for FGK (Solar-like) stars. Figure and caption adapted from Lawson et al. (2012), Mawet et al. (2012), and Dalcanton et al (2016).

Despite its 6.5m aperture JWST will be limited in angular resolution because it will operate at NIR wavelengths. Also it does not have a spectrograph fed by a coronagraph.

Even in the H-band, as plotted in figure 1, E-ELT and TMT will have excellent limiting angular resolution because of their large apertures. The excellent contrast prediction is due to the proposed combination of extreme AO and a coronagraph. The contrast performance at 763nm will not be as good as it is in the H-band.

LUVOIR(HDST)/HABEX are US proposals for a large space telescope with a high performance coronagraph. These are flagship mission concepts – most likely one of these will become the successor to JWST. The HZ Earth-sized planet biosignature search is the top priority science case for these mission studies which will be presented to the 2020 decadal review. Current estimates (which are highly uncertain) are for a launch around 2035. However, it should be noted that NASA's previous two flagship space telescopes (HST and JWST) both suffered long launch delays.

## Spectroscopic Biosignatures

The holy grail of the next era of exoplanet characterization will be to detect a robust biosignature gas combination in the atmospheres of terrestrial exoplanets. On Earth the most abundant biosignature gas is $O_2$, making up 21% of the modern Earth atmosphere. Ozone, the photochemical by-product of $O_2$ is also considered a biosignature. Both $O_2$ and $O_3$ have the potential for false positives, particularly for planets that are in a high UV environment from their host star such as around M dwarfs (Domagal-Goldman et al., 2014, Tian et al., 2014, Luger & Barnes 2014, Rameriz et al., 2014), which remain active for a much longer fraction of their main sequence lifetime than FGK stars. Most of the false positives for $O_2$ can be ruled out with an indicator of the reducing power of the atmosphere by detecting $CH_4$ (Harman et al., 2015; Domagal-Goldman et al. 2014). SUPERSHARP could measure $CH_4$ at $\lambda < 1000$nm and the E-ELT could measure it in the NIR. Thus, as first proposed by Lovelock (1965), the strongest biosignature still is the combination of an oxidizing gas ($O_2/O_3$) in combination with a reducing gas like $CH_4$ in a planetary atmosphere. Another false positive for $O_2$ could be due to the absence of a cold-trap followed by water photolysis (Wordsworth and Pierrehumbert 2014). An estimate of atmospheric pressure through detecting the $N_2$-$N_2$ or $O_2$-$O_2$ dimers would help identify this situation (Misra et al. 2014; Schwieterman et al. 2015). The $O_2$-$O_2$ dimer has features at 1.06$\mu$m and 1.27$\mu$m which could be observed with E-ELT. Other biosignatures include $N_2O$, $CH_3Cl$, and DMS though these gases are present in too low of concentration at Earth-level concentrations to be detected with the next generation telescopes unless a very low-UV environment allows for their build-up in a planetary atmosphere (Grenfell et al., 2012, Rugheimer et al., 2015).

Most of the known false positive mechanisms for $O_2$ and $O_3$ are informed in part by the UV environment of the host star (Hu et al., 2012; Domagal-Goldman et al., 2014; Luger and Barnes 2015; Harman et al., 2015). As well, the ratio of far-UV to near-UV is what determines the abundance of $O_3$ in an exo-Earth atmosphere (Segura et al., 2005). Thus it will be vital to have contemporaneous measurements of the UV stellar radiation field of habitable exoplanets in order to contextualize the atmospheric abundance of biosignatures in habitable exoplanets. Since SUPERSHARP's proposed wavelength reaches into the UV down to 120nm, UV observations of the host-star will help inform our interpretations of an $O_2$ detection.

Even with the potential false positives for $O_2$, detecting oxygen in an exoplanet atmosphere will be a vital technological step forward for understanding the frequency of both potential biosignatures and false positives in the universe. Detecting $O_2$ in a planet on the inner edge of the habitable zone that could be in a run-away greenhouse state or in a planet orbiting a pre-main sequence M dwarf will not constitute a detection of life, but it will help our understanding of the frequency and potential parameter space of $O_2$ and will provide context for more plausible biosignature detections.

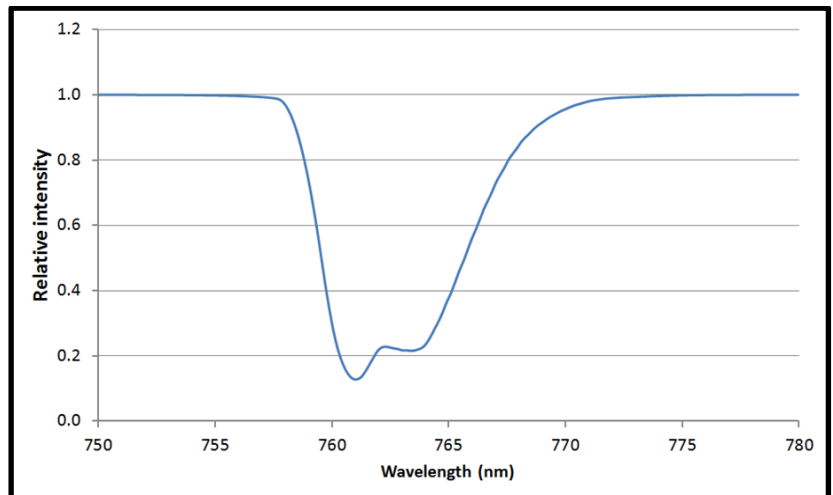

Figure 2. The Oxygen A-band biosignature at low resolution.

While transit spectroscopy is a vital tool to characterize the atmospheres of habitable planets, particularly around M dwarfs, ultimately a direct detection mission will be needed due to the transit geometry allowing for only a small fraction of potential targets to be characterized. Since transits are infrequent for Earth-like planets in the HZ, such planets are more amenable to characterization with a direct detection mission. Additionally, for an Earth-Sun system, refraction effects limit the depth of the atmosphere probed to 12 km and higher even in absence of clouds or hazes with transmission spectroscopy (Bétrémieux & Kaltenegger 2014).

## The habitable zone

We have used the definition of the habitable zone given by Kopparapu et al 2013 for our yield calculations (section 5, page 15). There are other published versions but the choice of HZ does not affect the calculated yield significantly (the width is the important parameter).

## Resolving the HZ for nearby host stars

We have a good knowledge of the stellar content of our local neighbourhood and our knowledge of the occurrence rates of planets (Fressin et al, 2013, Dressing and Charbonneau, 2015, Petigura, et al, 2013) indicates that *all* of these stars are likely to host exoplanets (habitable or not). Using the simplifying assumption that 10% of stars host a habitable planet (i.e. eta-Earth=0.1) and our knowledge of the size of the habitable zone we can do some simple calculations to estimate limits on the angular separation required to search for biosignatures in a sample of ~100 habitable Earths. Here we are ignoring contrast and exposure time so the telescope sizes calculated are lower limits. A full analysis that includes predicted exposure times and measured occurrence rates is described later in section 5 (page 15).

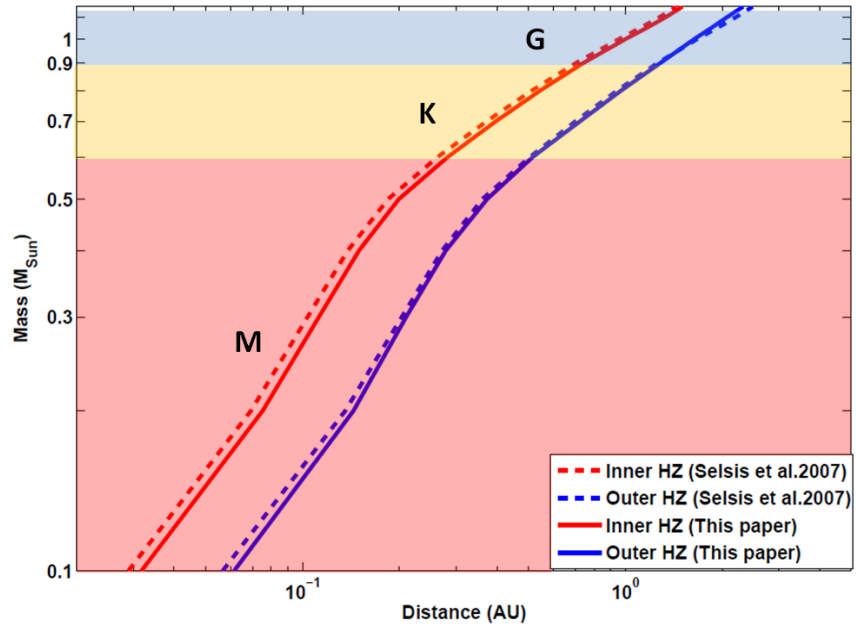

Figure 3. The extent of the habitable zone as a function of stellar mass and spectral types, G, K and M (Figure adapted from Kopparapu et al 2013).

| host star type | N subsample size | dist. lim. for N stars (pc) | host I mag limit | HZ size AU | planet-star contrast | liming ang. sep. mas | planet I mag limit | telescope aperture size (m) |
|---|---|---|---|---|---|---|---|---|
| F stars | 85.2 | 43.5 | 6.1 | 2.80 | 2.13E-11 | 64 | 32.8 | 8.8 |
| G stars | 8.4 | 15.5 | 5.2 | 1.00 | 1.67E-10 | 64 | 29.6 | 8.8 |
| K stars | 6.3 | 9.3 | 5.8 | 0.60 | 4.63E-10 | 64 | 29.2 | 8.8 |
| early M stars | 0.3 | 3.1 | 5.2 | 0.20 | 4.17E-09 | 64 | 26.1 | 8.8 |
| late M stars | 0.0 | 1.2 | 6.0 | 0.08 | 2.60E-08 | 64 | 24.9 | 8.8 |

Table 1: This table bins the local main sequence stars into 5 groups and shows the contrast, the angular separation limit and the magnitudes of the host and planet at the liming distance of the subsample. The 5 groups combined provide a total sample of ~100 HZ Earth-sized targets (see text for details). In this example the number in each bin was chosen so that the telescope aperture was the same in each case which is the smallest aperture that can actually resolve 100 HZs. The contrast and angular separation are plotted in figure 1. The planet's I band magnitude limit assumes maximum elongation and an albedo of 30%.

In table 1 nearby FGKM main sequence stars (binaries and multiples are excluded) are binned into 5 groups. The second column shows the chosen subsample size (N) which is the number of stars that actually have an Earth-sized planet in the HZ (which assumes eta-Earth=10%). The third column shows the limiting distance one needs in order to have N stars in the subsample. Using the mean HZ size for the group and a planet radius of $1 \times 10^7$ metres (1.57 $R_\oplus$) we can then calculate the planet-star contrast and the angular separation at the distance limit.

$$Contrast = \frac{albedo}{8} \times \left( \frac{planet\_radius}{orbit\_radius} \right)^2$$

In table 1 the subsample sizes were chosen to give the same limiting angular separation $\theta$ for each group and also chosen so that the total sample comes to 100 targets. The resulting telescope size of 8.8m for the 763nm $O_2$ biosignature (= 3 × 1.2 $\lambda/\theta$) therefore represents the absolute minimum telescope size that can resolve 100 HZ Earth-sized planets (assuming eta-Earth=0.1). The distribution of targets is dominated by F star hosts because these are the most easily resolved. However, in practice these are hard to observe because of their difficult contrast.

Table 2 is similar to table 1 except now the sample size of 100 excludes the F-stars. In this case a telescope size of 16.5m is needed and most of the sample is made up of K and G stars. Clearly therefore, an important design aim is to maximise the telescope expansion factor (i.e. the ratio of its deployed size to its folded size) simply to have enough targets outside the IWA.

These simple calculations show that we need a large telescope to have any chance of having a good sized sample (~100). However, the number of potential targets grows as $D^3$ (where $D$ is telescope aperture or baseline) so once we are in the regime where we have a few targets it only takes a small increase in telescope size to get a lot more targets. A 26% increase in telescope size will double the number of targets whereas a 215% increase will increase the number of targets by 10-fold. It's all about the baseline. **Primary mirror baseline is more important than collecting area and we must consider unfilled primary mirror geometries.** In practice the primary mirror collecting area will be limited by both the mass budget and the financial budget and the expansion factor for the deployment of the telescope structure will determine the primary mirror baseline and therefore the yield.

| host star type | N subsample size | dist. lim. for N stars (pc) | host I mag limit | HZ size AU | planet-star contrast | liming ang. sep. mas | planet I mag limit | telescope aperture size (m) |
|---|---|---|---|---|---|---|---|---|
| | | | | | | | | |
| F stars | 565.0 | 81.7 | 7.5 | 2.80 | 2.13E-11 | 34 | 34.1 | 16.5 |
| G stars | 55.6 | 29.2 | 6.5 | 1.00 | 1.67E-10 | 34 | 31.0 | 16.5 |
| K stars | 42.1 | 17.5 | 7.2 | 0.60 | 4.63E-10 | 34 | 30.6 | 16.5 |
| early M stars | 2.2 | 5.8 | 6.5 | 0.20 | 4.17E-09 | 34 | 27.5 | 16.5 |
| late M stars | 0.1 | 2.3 | 7.3 | 0.08 | 2.60E-08 | 34 | 26.3 | 16.5 |

Table 2: As for table 1 but here the 4 lower groups (GKM) combined provide a total sample of ~100 HZ Earth-sized targets. Note that a couple of early M stars have crept in to the sample. It also predicts zero late M-stars. The recent announcement of a habitable planet orbiting Proxima Cen appears to show that we are lucky to have such a nearby target to study.

Here we have used a yield of 100 to estimate telescope size. Later we will present some results from our full yield calculator where the detailed telescope geometry and contrast performance is used to calculate yield. We will see that our designs with a baseline of 24m (set by the launcher size) and a raw contrast of $10^{-8}$ fall a little short of our goal of a yield of 100.

## Expected frequencies of planets

We now know (mostly from the analysis of Kepler data in the last 3-4 years) how many planets a star is likely to have. We can use this information to more precisely determine what is needed to observe ~100 HZ Earth-sized planets and our yield calculator is described in section 5. Figures 4, 5 and 6 show examples of planet occurrence data that can be found in the literature. We now know that planets are relatively common (this was not known when the pioneering biosignature mission design studies such as Darwin and TPF were first carried out).

## Planets seen via their reflected light

Figure 7 shows how the brightness of a planet (seen in reflected light) relative to its host star (planet-star contrast) depends on orbital radius. Although similar to figure 1 this plot does not depend on any observational parameters such as the star's distance or the telescope size (ignoring the scale along the top of course).

To minimise the inner working angle (IWA) we have to maximise telescope baseline and minimise $\lambda$ (IWA = 3 × 1.2 $\lambda$/[mirror baseline] = Orbital Radius in radians). The planet-star contrast that can actually be observed depends on the WFE. Detailed end-to-end modelling of the performance of a particular telescope design allows us to predict and therefore minimise the WFE. In practice, we aim to maximise the light received from the planet and minimise the light from the parent star that ends up in the same pixels as the planet's light whilst also making a control measurement so that the parent star's contamination can be accurately determined and subtracted.

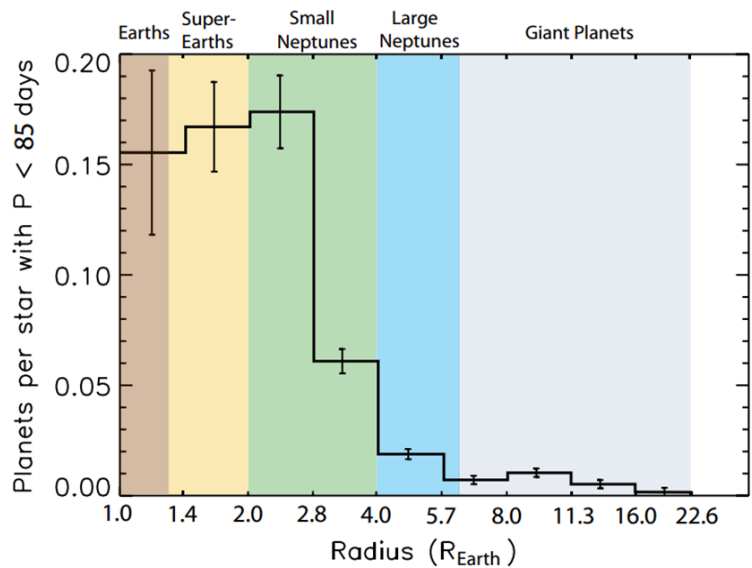

Figure 4: Average number (corrected) of planets per size bin for main sequence FGKM stars, determined from Kepler data (Figure from Fressin et al 2013).

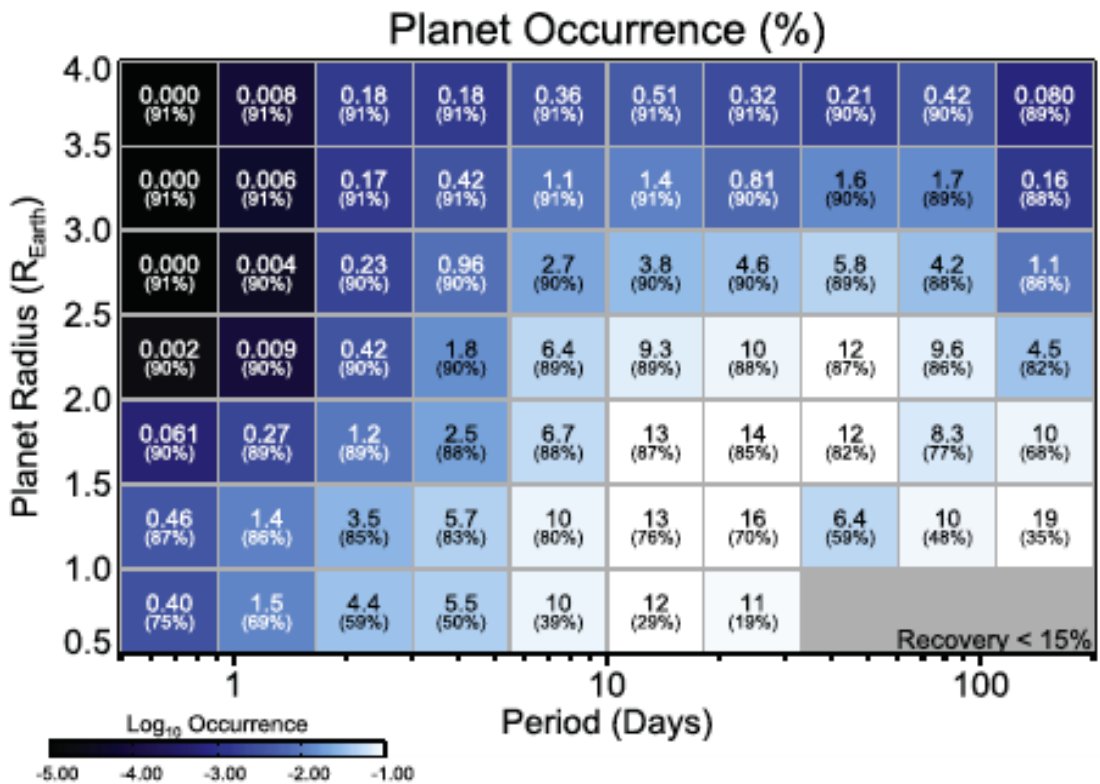

Figure 5: Planet occurrence rates for M stars (Figure from Dressing and Charbonneau 2015).

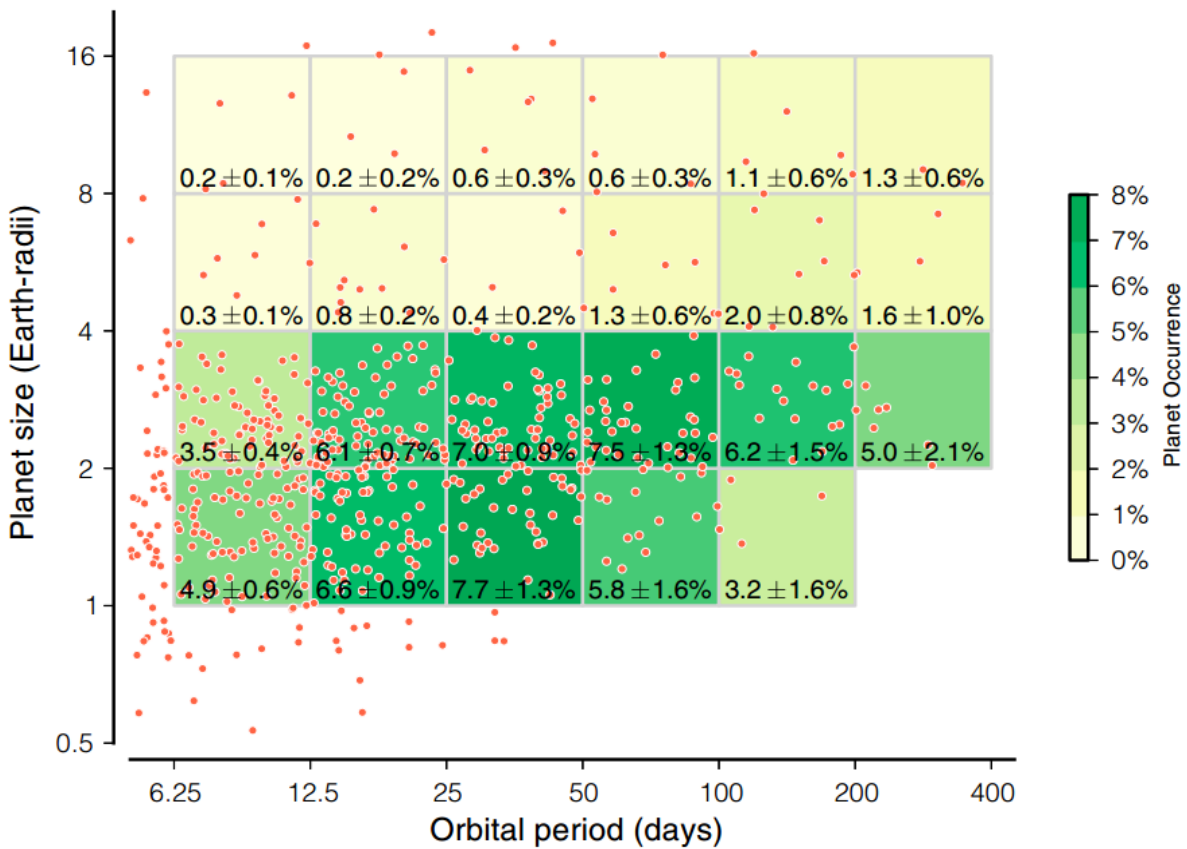

Figure 6: Planet occurrence rates for FGK (Sun-like) stars (Figure from Petigura et al 2013).

# 3) Additional exoplanet science

Figure 7 covers a large range of possible planet sizes and orbital radii and so captures most of the diversity we wish to explore in exoplanet research in general. Our objective is therefore to be able to observe over as much of this plane as possible. Of course, planets that are larger than the Earth and self-luminous planets will be easier to observe than the biosignature targets described above. Nevertheless, these are also extremely interesting scientifically for the understanding of the structure, formation and evolution of exoplanets.

## Architectures of planetary systems

With direct imaging we can see all the exoplanets above the contrast limit on orbits large enough to be spatially resolved. We will therefore be able to get the architecture of the planetary system of most nearby stars. If we make multiple visits this capability is independent of orbital inclination (unlike transits and RVs) and it will be common to find multiple planets and unusual to find nothing at all. SUPERSHARP will explore Mars-size to Jupiter-size planets on orbits from 0.03 to 100AU orbiting nearby M to A type stars. For all orbital inclinations, SUPERSHARP will give us the planetary system architecture for every star it observes and for each planet it will provide time-dependant astrometric, photometric and spectroscopic information in the wavelength range 150nm - 1000nm, revolutionizing our knowledge of planetary systems.

## Spectroscopic characterisation of exoplanets in general

To characterise an exoplanet we obtain its spectrum and then apply powerful modern retrieval methods. These spectra can be from transit spectroscopy or as in the case of SUPERSHARP from direct imaging/spectroscopy. The retrieved atmospheric model tells us the gas species abundances, the temperature and pressure profile, the atmospheric composition, the clouds and hazes and $\log(g)$.

By spectrally characterising many exoplanets we can do comparative planetology which will greatly help us understand planet formation and evolution. For a single star, the region in figure 7 that allows a biosignature search is very small (small planet in the HZ). However, we know that generally speaking exoplanets are very common and so we expect to find several other planets in other locations in the diagram (like the 7 solar system planets plotted in addition to the Earth). Planets on orbits bigger than 0.4AU will be relatively easy to characterise especially the larger ones. This is highly complementary to transit spectroscopy which is mostly sensitive to planets on small orbits (<0.4AU) and adversely affected by clouds and haze. Thus spectral characterisation and comparative planetology with SUPERSHARP will probe most of the diversity of the exoplanet population.

## Proxima b

The discovery of a planet with $M\sin i=1.27$ $M_\oplus$ orbiting Proxima Centauri was recently announced (Anglada-Escudé et al, 2016). This is an extremely exciting result because the planet is Earth-sized, in the HZ and it orbits our Sun's nearest stellar neighbour. Getting its spectrum cannot be done by transit spectroscopy because its orbit is not edge-on. Its 0.05AU orbit projects to $\theta$=38mas on the sky so at 763nm and assuming the IWA is 3 Airy disks across then this corresponds to D = 3×1.2$\lambda$/$\theta$ = 14.6m which makes it difficult for the proposed 12m HDST (see figure 1) even though it is our Sun's nearest stellar neighbour. On the other hand the 24m versions of SUPERSHARP can easily observe this object because it is outside the IWA and its contrast is ~$10^{-6}$ – $10^{-7}$. The E-ELT will certainly be able to do imaging and spectroscopy of this in the NIR. Whether it will be able to do this for the 763nm biosignature is not clear – it depends on how well the AO systems work at this wavelength. This is a good example of why maximising telescope baseline (aperture) is extremely important.

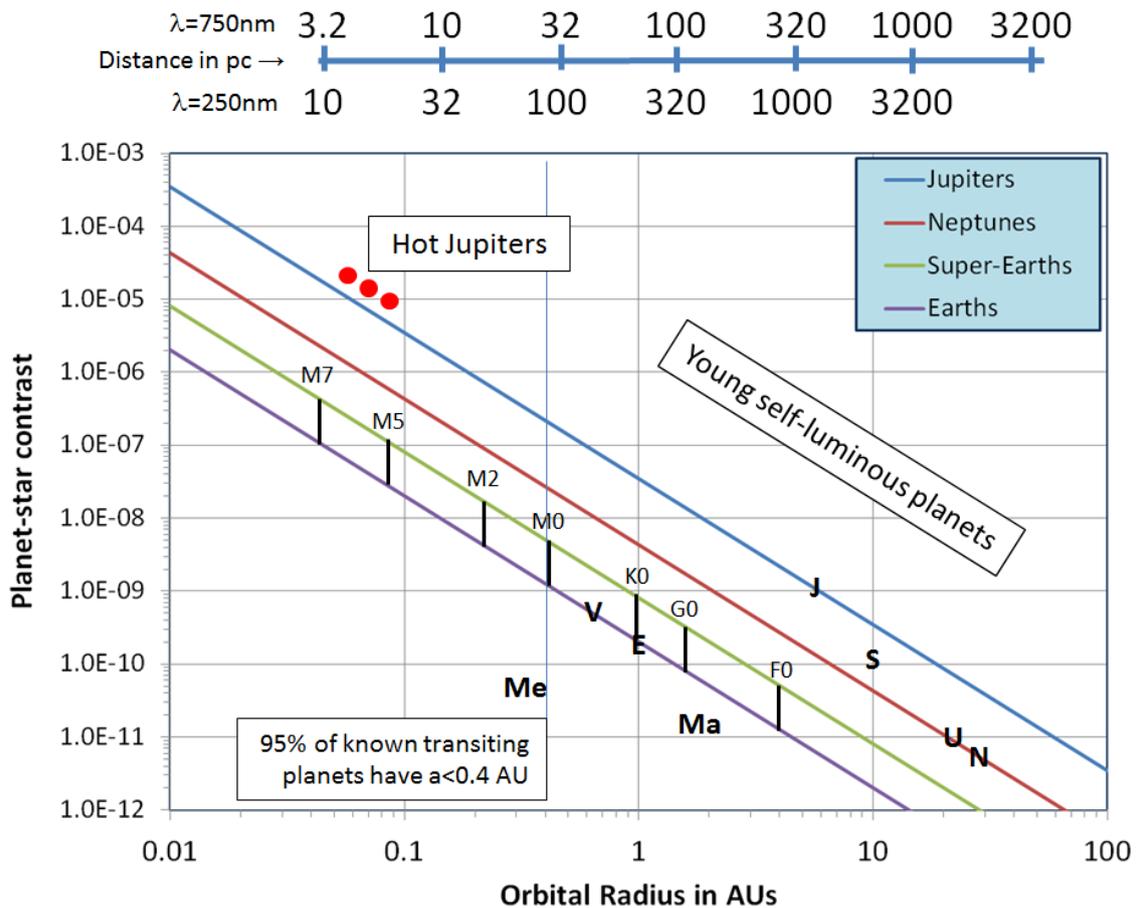

**Figure 7:** Planet-Star Contrast versus orbital size for exoplanets. Circular orbits are assumed and the contrast is for maximum angular separation. The diagonal lines are for exoplanets seen in reflected light. The scale along the top shows the distance from the solar system at which the orbital radius equals the IWA for a primary mirror baseline of 24m and for 250nm and 750nm ($O_2$). The mean size of the habitable zone is also plotted as short vertical lines for host star spectral type. Habitable planets orbiting F-stars are difficult because of their contrast. Habitable planets orbiting M-stars are difficult because of their apparent separation. The eight planets in our solar system are plotted using their initial letters. Note that 95% of known *transiting* planets are to the left of the vertical line at 0.4AU whereas SUPERSHARP studies planets with orbits > 0.03AU. The two techniques are highly complementary.

## Direct imaging of Hot Jupiters

A 24m version of SUPERSHARP operating at 121nm (Lyman-alpha) will be able to image hot Jupiters. For example 51 Peg b has a separation of 3.4mas and $1.2 \times 121nm/24m$ gives a resolution of 1.25mas so it can be resolved. Note that the contrast for hot Jupiters is relatively easy at ${\sim}10^{-5}$. This is another example of how science is enabled by a large telescope baseline. Ups And is slightly wider at 4.4mas. A 24m telescope should be able to resolve ${\sim}10$ nearby hot Jupiters (which would be studied independently of the main biosignature sample).

# 4) Scientific Requirements

## Sample size

A sample of ${\sim}100$ habitable-Earth containing systems should be assumed as the goal for the purpose of designing the mission. This is to allow for the uncertainty in the planet occurrence rates and to ensure that a null result still provides a useful constraint on the frequency of life-bearing planets (assuming $O_2$ false positives can be ruled in or out by checking for methane or dimers or UV observations). See Dalcanton et al (2015) for further details.

## Sample selection

To do the search we need to know beforehand which stars actually have HZ Earths. There are several possible ways we can get this information.

1) Will we know some of this from other facilities by 2039. e.g. HARPS3.
2) We could use a smaller direct imaging mission to act as a finder. e.g. to match the IWA at 763nm of a 20m telescope we could use a 4m telescope in the UV. This could also act as a technology demonstrator for SUPERSHARP.
3) We could use the large SUPERSHARP telescope itself in survey mode for the first year of the mission (imaging is faster than spectroscopy).

## Mission duration

We propose that SUPERSHARP be dedicated to exoplanets only and that a 5 - 7 year mission duration be adopted. This comes from the longer exposure times needed to work with an instrument contrast of ${\sim}10^{-8}$ and the need to do a finding survey at the start of the mission.

## Wavelength and spectral resolution

The best biosignature is the $O_2$ feature at 763nm and it can be observed with a spectral resolution of $R{\sim}100$. A relatively small total wavelength coverage of only 30nm is needed for the biosignature itself (see figure 2) but more wavelength coverage would help with speckle subtraction. Other biosignatures or spectral features to discriminate against false positives may be included in the specifications. The UV offers the best spatial resolution and contemporaneous UV measurements are important for ruling out biosignature false positives. The UV should therefore be included in the mission specification at low spectral resolution (R~20).

# 5) Measurements concept (technical requirements and mission concept)

## Yield calculator

We have built a calculator which tells us how many biosignature targets (the yield) can be observed for a specific telescope design and mission duration. This is an essential tool for developing the technical requirements and the mission concept. The calculator uses the full census of the local star population (out to 85pc) and randomly assigns HZ Earths to the main sequence stars using the observed occurrence rates from Kepler (with some extrapolation for large HZ orbits). Monte Carlo modelling is also used for the position of the planet within the HZ and the planet's radius. The exposure times required to detect the 763nm $O_2$ biosignature for these are then calculated (using an assumption about the speckle background brightness) and they are sorted by exposure time. The yield is derived by working down the list until the total exposure time is equal to the time available in a 5 year mission.

| SUPER-SHARP | Primary mirror | Effect. Diam. | Base-line | yield | IWA (mas) |
|---|---|---|---|---|---|
| Big cross | 4 x 10 x 2.8 | 11.9m | 24m | 71 | 16 |
| Big strip | 2 x 10 x 3.4 | 9.0m | 24m | 58 | 16 |
| Small strip | 2 x 7.5 x 1.8 | 5.8m | 15m | 26 | 25 |

**Table 3:** The three versions of SUPERSHARP that were used as inputs to the yield calculator. The IWA refers to 763nm. These fall short of the goal of a yield of 100 because an instrument contrast of $10^{-8}$ was assumed (compared with $10^{-10}$ or $10^{-11}$ for the current NASA studies).

We have used the yield calculator for 3 SUPERSHARP designs which are listed in table 3 along with the yield result. Illustrations of the big strip and the big cross telescopes are shown in figures 13 and 10. Our yield results are also shown in figure 8 in comparison to the results of Stark et al, 2015. SUPERSHARP's non-circular geometry improves the yield in the sense that there are more targets per unit area of primary mirror than for a circular geometry.

## End-to-end simulations

We have also done full end-to-end (telescope + coronagraph) wavefront propagation modelling of SUPERSHARP using John Krist's PROPER IDL library. Figure 9 shows one result for a 3rd mag G-star host at a distance of 5pc with an exoplanet companion of 25.5mag (i.e. a contrast of $1 \times 10^{-9}$). The exposure time=500hr and the spectral resolution is R=100 at 750nm. ADI subtraction is assumed and the telescope modelled was the 24m big cross design. Our next step will be to use the end-to-end simulations to derive the instrument contrast for our yield calculator.

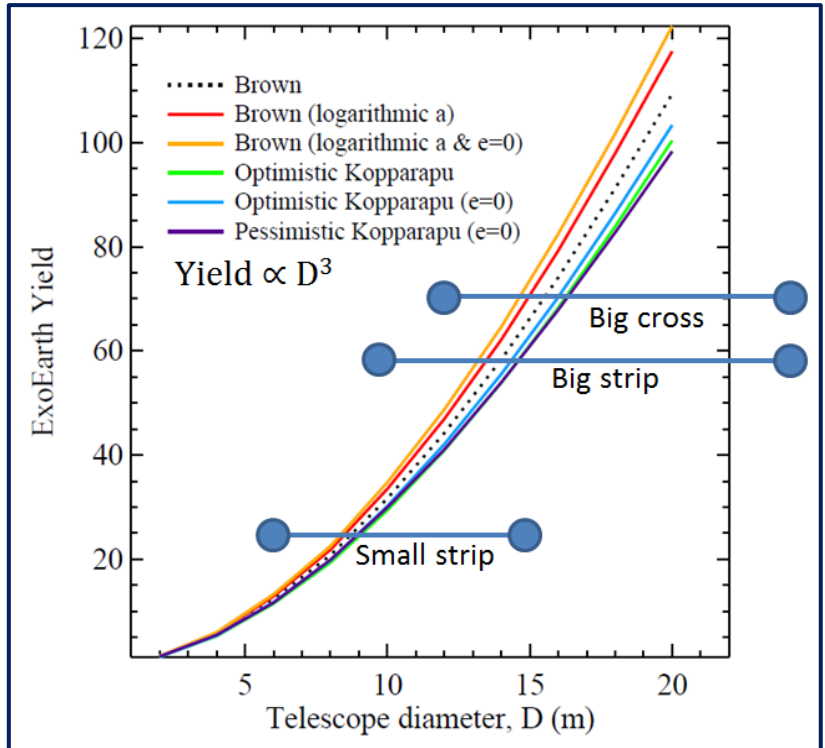

**Figure 8:** Yield versus telescope size. The underlying figure was taken from Stark et al, (2015) and shows different HZs. Over-plotted are the yield values for our 3 SUPERSHARP models (see table 3). They are plotted as horizontal lines because the mirror geometry is not circular. The left end of the line corresponds to the diameter of a mirror with an equivalent collecting area. The right end of the line corresponds to the maximum baseline of the mirror. In all 3 cases SUPERSHARP's non-circular geometry delivers a better yield per mirror collecting area than a simple circular mirror.

## IWA

Earlier (page 8) we showed that a minimum IWA of 64 mas is needed to see the 100 most easily resolved HZs assuming eta-Earth=10%. For this to work the mission duration must provide enough time to actually observe all 100 targets. In practice, this will not be the case because the F-stars take too long to observe as shown by our yield calculator which gives a better estimate of the required IWA (~16mas to get 71 targets). An important point is that IWA is directly related to the sample volume whereas the limiting contrast is related to the exposure time for a target that lies outside the IWA. This means that increasing telescope baseline (decreasing IWA) allows access to more targets whereas improving limiting contrast only increases the yield for time-limited surveys – once all of the targets outside the IWA are done the yield stops growing. It is therefore important to maximise the telescope baseline that can fit in the launcher's fairing (i.e. maximise the expansion factor). Some resultant increase in the WFE can be tolerated but of course the expansion factor must not put the design over-budget.

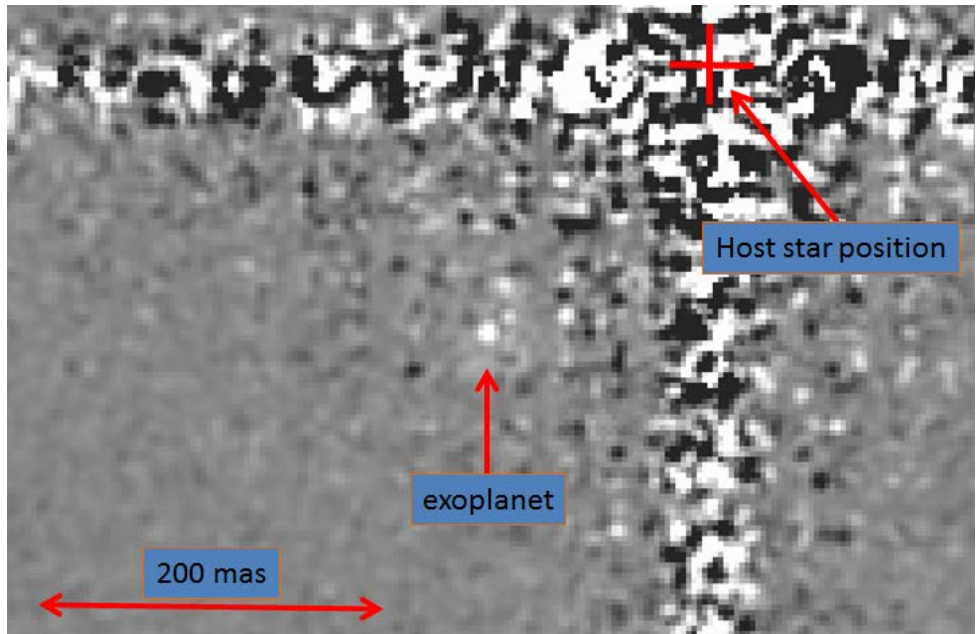

**Figure 9:** Simulation of a single channel in the data cube (7.5nm bandpass, after speckle background subtraction) of a HZ super-Earth orbiting a G star at 5pc distance from the Earth. The planet can be clearly seen against the noise field. The integration time is 500 hours and the image scale is 3.2 mas/pixel. The planet image can be distinguished from the planet-like speckles because the speckles move with wavelength and so appear in different positions in other slices of the data cube while the planet image stays still.

## Telescope size

For the biosignature search we need to be able to spatially resolve the physical size of the HZ for ~100 stars that actually have an exo-Earth orbiting in the HZ. For the $O_2$ biosignature at 763nm and an IWA of 16mas this corresponds to a baseline of 24m as per the yield calculations.

## Exposure times

We need to get good signal-to-noise for a target which is only a few 10s of mas away from a star which is 1 million to 10 billion times brighter ($15 - 25$ mags). We therefore need to reduce the amount of light from the parent star at the location of the planet. Let $T$ be the exposure time, $Z$ the required signal-to-noise ratio and $R_P$ and $R_B$ the photon arrival rates from the planet and the background (from the adjacent host star). Then it can be shown that

$$T = \frac{Z^2 \left(R_P + 2R_B\right)}{R_P^{\,2}} \approx \frac{2Z^2 R_B}{R_P^{\,2}}$$

For an acceptable exposure time $T$ this tells us what maximum value of $R_B$ we can tolerate. This equation assumes there are no systematic errors and the performance is photon-noise limited. This equation was used in the yield calculator assuming $R_B/R_{star} = 10^{-8}$ (which is a less ambitious number than is typically used by the NASA teams).

It is essential to eliminate (or at least drastically reduce) systematic errors when subtracting off the background light. Let's assume that $R_{BE}$ is the estimated background rate and $R_{BA}$ is the actual background rate that was mixed in with the light from the planet. Also let's assume that the estimated rate is slight wrong so that $R_{BE} = f\, R_{BA}$ where $f \sim 1$ but is not equal to one. In this case the final signal-to-noise obtained as $T$ goes to infinity is given by

$$Z = \frac{R_P}{(1 - f)\, R_{BA}}$$

So we see that for significant systematic errors we reach a point where further integration does not improve the value of Z. For example for $f$=0.99 and $R_{BA}/R_P$=100 then $Z$=1! In practice, if we are trying to measure a planet signal which is 1000 times fainter than the speckle flux it is mixed up with then we need $1 - f \sim 1 \times 10^{-5}$. Coronagraphic instruments are typically limited by systematic errors.

## Description of SUPERSHARP

The description given here refers to the "big cross" concept which is capable of addressing the science case given above. Figure 10 shows the "big cross" version of the SUPERSHARP idea.

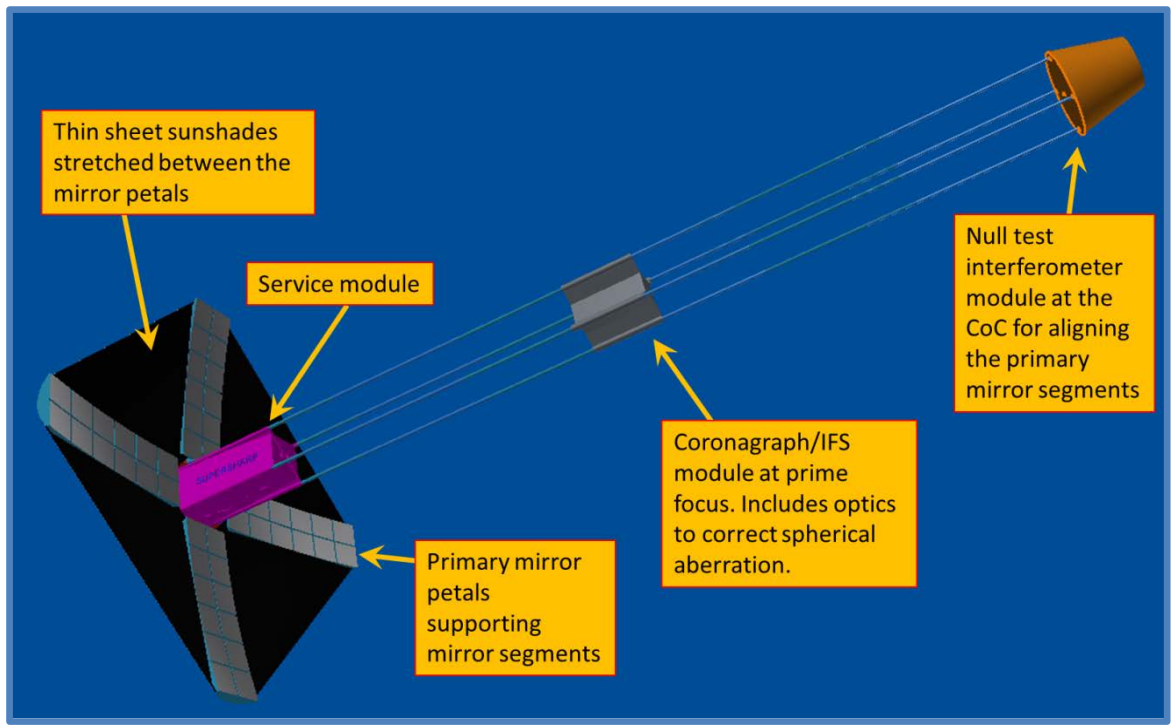

**Figure 10:** Image of the "big cross" version of SUPERSHARP. When folded this fits in to an Ariane 6 launch vehicle. When deployed, this version of the SUPERSHARP concept is 52m long and 19m x 19m wide. The primary mirror baseline is 24m.

**Primary mirror geometry:** Our yield calculator shows that an unfilled aperture can give the same yield (number of observable biosignature targets) as a filled aperture but with less mirror area (by increasing the baseline). This will reduce mass and cost. We have so far made CAD models of two geometries: a long, one dimensional strip 24×3m (see figure 13) and the 24m cross shape that is described here (see figure 10).

**Primary mirror segments:** The mirror segments will have an rms surface accuracy of 5-30nm (for comparison, the rms accuracy of the 1.5×0.5m GAIA mirrors is 9nm). To stay within the launch mass budget the total mass per unit area of the primary mirror segments must be less than 20 kg m$^{-2}$. This is consistent with the SiC mirror technology used for Herschel and Gaia which can now achieve a mirror surface density of 16 kg m$^{-2}$. Spherical mirrors are the easiest to manufacture and are the same for all segments. A spherical primary is also the easiest to measure with a laser interferometer. For spherical segments lateral (xy) offsets and rotational offsets have no effect on the image quality. Each mirror can therefore be positioned using 3 (rather than 6) actuators. However, the telescope has very bad spherical aberration at the prime focus and so a prime focus corrector is needed. For exoplanets, only a small FOV is needed and this is feasible with a corrector. The SUPERSHARP concept can still deliver a large FOV if required for other applications (see section 6). In this case an aspheric primary must be used which means more actuators and more segment types.

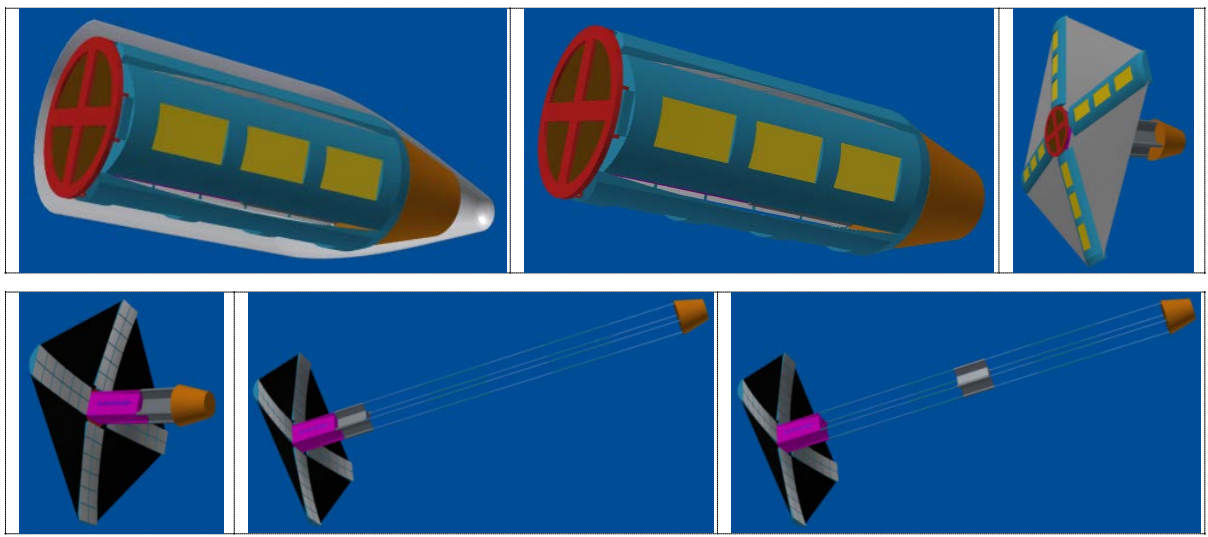

**Figure 11:** From top left to bottom right, a time sequence of images from being fully folded inside the Ariane 6 fairing to being a fully deployed telescope. After the rocket fairing is jettisoned the telescope is about 4.4m in diameter by 10m long. Next the four mirror petals unfold outwards and simple fabric sunshades unfurl between the mirror petals. There are solar panels on the back of the mirror petals. The four support tubes extend to their full length in a way similar to the deployment of an electric car aerial. This puts the null-test interferometer at the centre of curvature. Finally, the instrument module housing the telescope's corrector optics, the coronagraph and integral field spectrometer runs along the extended tube structure to the required position for focusing on objects at infinity.

**Deployment (unfolding):** The whole telescope must unfold from approximately 4.5m diameter × 10m when inside the Ariane 6.4 fairing to its operational configuration which is perhaps as large as 24m × 24m × 52m. This is an expansion factor of ~5. We have developed a simple working concept (to demonstrate basic feasibility) for the folding and deployment. Figure 11 shows the deployment sequence.

**Maintaining telescope WFE quality:** This large light-weight telescope structure will inevitably be prone to dimensional drifting which must be compensated for by active/adaptive control of the optics. A laser null test interferometer continuously measures the alignment of the segmented primary with a few nm precision at least once every second. Figure 12 shows schematically how this works. Note that it can be used for both spherical and aspheric primary mirrors although for SUPERSHARP a spherical primary is envisioned. The laser wavelength is chosen to be one which is not of scientific interest and is filtered out by the science instruments. When the mirror alignment has drifted out of specification the mirror segment actuators are adjusted to return the mirror back to correct alignment. The shutter is closed during this adjustment to wait for acoustic effects to damp out. The drift time therefore has to be at least 10 times longer than the acoustic damping time. The silent, micro-gravity environment of space is an advantage. For JWST the engineers put a lot of investment into making the system very stable with a drift time of about 2 weeks but even a drift time of only a few minutes should be acceptable for SUPERSHARP reducing the costs and associated risks of any sun-shields and thermal management systems.

**Instruments:** The optical train will include one or more coronagraphs and integral field spectrographs. Further WFE control will be applied just before the coronagraphs via deformable mirrors. The FOV will be ~1000 × 1000 mas and the spectral resolution will be R~100. Only one UV detector and one CCD are required for the science data acquisition and the data download rate is relatively low compared to many existing and planned missions.

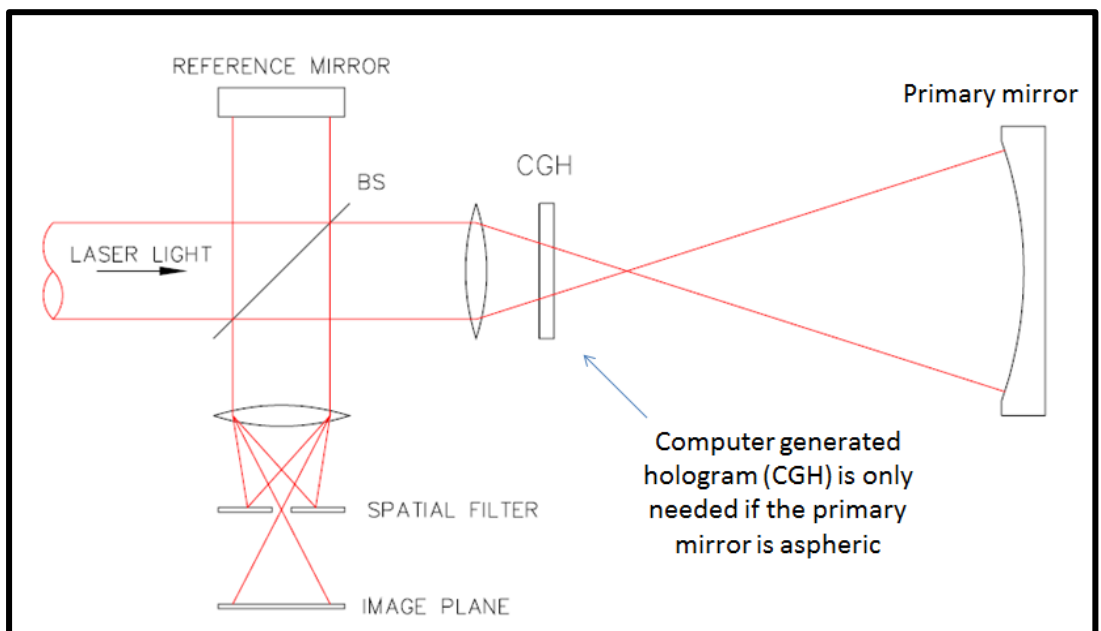

**Figure 12**: Using a null-test interferometer to test a concave mirror. This is a very commonly used technique for manufacturing optics and can achieve very high surface accuracy. The image plane shows fringes which correspond to surface errors on the primary mirror. When the mirror segments are all perfectly aligned the fringes disappear.

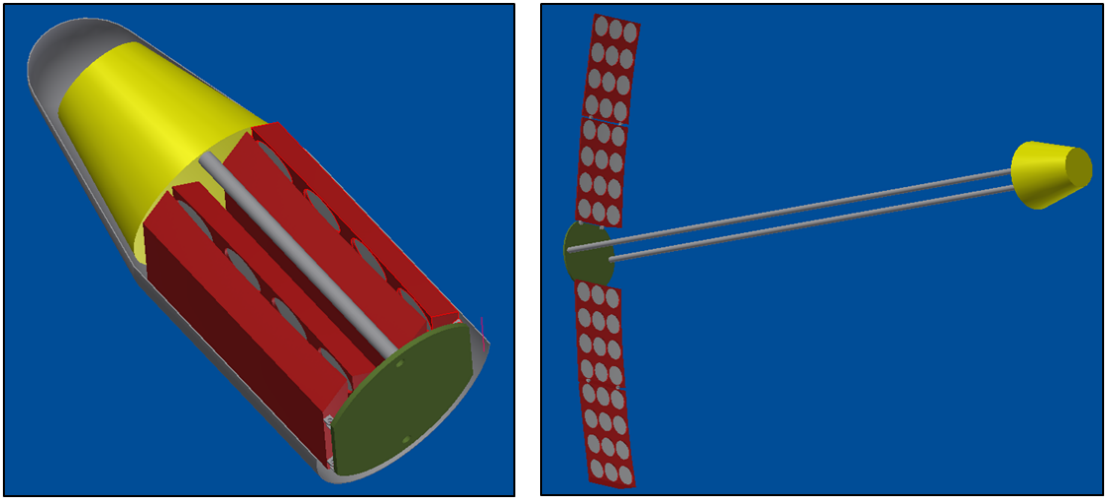

**Figure 13:** Image of the "big strip" version of SUPERSHARP. This was an early version which was designed to fit into a Soyuz launch vehicle. It does not have an interferometer at the centre of curvature. The image on the left shows how it folds into the Soyuz fairing. The image on the right shows it deployed. The primary mirror is 24m × 3.4m and the telescope structure is 30m long.

## How can we make SUPERSHARP affordable?

In the US it is widely assumed that the LUVOIR/HDST concept will require a budget similar to those for HST and JWST (~US$10B) whereas the HABEX concept assumes a budget of up to ~US$5B. Of course at this stage these are concept studies so no firm costs are available. However, clearly these both seem to be substantially more than the cost of an L-class mission (~€1.0B total cost to ESA). Given that SUPERSHARP shares the same biosignature HZ Earth-sized science goal how can it possibly fit within an L-class budget? Here we list some key recommendations on how to make SUPERSHARP less expensive than the US concepts.

- **Relax the instrument contrast requirement**: The US studies argue that a speckle contrast of $10^{-10}$ is needed but recent ground based observations suggest that this can be relaxed by ~100-1000× by accepting longer exposure times and eliminating systematic errors. See the section above on systematic errors.
- **Abandon the circular mirror:** The primary mirror has to have at least one large baseline to minimise IWA but it does not have to be a filled circle. Baseline is more important than diameter/area. Reducing fill factor significantly reduces costs. If the fill factor is reduced by increasing baseline this does not necessarily increase exposure times because the PSF of the planet image gets smaller and so the underlying speckle background gets fainter.
- **Use spherical primary segments.** These need fewer actuators and they are easier and less expensive to make.
- **Continuously and rapidly align the primary segments:** This relaxes the requirements on the telescope stability (so we don't need a complicated thermal management system or a separate sun shield).
- **Dedicate the mission only to exoplanets:** SUPERSHARP should not be general purpose so it should not have instruments and design features that cater for a broad range of science goals.

- **Keep it simple wherever possible:** Wavelength range 110 − 980nm. Very small FOV (1.0 x 1.0 arcsec). Low spectral resolution of R~100. Relatively inexpensive coronagraphs feeding integral field spectrographs. Very few detectors. No IR detectors. No cryogens. Telescope not cooled. Reducing the number of subsystems obviously reduces costs because those subsystems no longer have to be paid for. But the cost saving is much greater than this in practice because it reduces the number of interfaces between subsystems and each of these interfaces is very expensive. For example, adding a subsystem to a system that already has 5 subsystems adds not only the cost of the new subsystem but also the cost of the 5 new interfaces that are required.

# 6) SUPERSHARP: beyond exoplanets

The SUPERSHARP concept, i.e. a folded, deployable, actively controlled space telescope, was created because of the strong scientific desire to set meaningful constraints on the frequency of life-bearing planets in the universe (the "new science" that we want to enable). However, SUPERSHARP is a potentially revolutionary technology:

1. For any fixed budget, SUPERSHARP essentially allows us to have a much bigger telescope than was previously thought possible. For ESA's cosmic vision program this applies to all three (S, M and L) mission budget envelopes. Many branches of observational astronomy can benefit from this.
2. Furthermore, as time goes on it may become harder to propose truly exciting new missions because the obvious ones have already been done and the funding levels for new missions will most probably stay the same for the next 20 years. The SUPERSHARP concept could therefore add significant scientific value to ESA's already excellent UV, optical and IR, space observatory program.
3. ESA's substantial Earth Observations (EO) program can also benefit too. A SUPERSHARP telescope can provide much higher ground definition from LEO than was previously possible or it can provide a major EO platform in geostationary orbit (which has been the subject of the two ESA design studies, GEO-OCULUS and "towards 1-m from GEO").
4. EO in general is a huge and rapidly expanding global market. There are about 330 telescopes in space right now looking down at the Earth − far more than there are looking up! Combining this with the micro-sat revolution (which is making all aspects of space hardware and launches more affordable) SUPERSHARP has great potential for commercial exploitation.

# 7) The SUPERSHARP roadmap

We foresee that the following steps will be required to eventually get to a full sized (~24m baseline) biosignature searching SUPERSHARP L4 mission.

1. Engage with ESA on a preliminary design study. The applicants will provide further effort on the end-to-end simulations, the yield calculator, the design of the telescope optics, the design and control of the coronagraph(s) and further development of the science case. ESA will work on the engineering implementation of the folded, deployable structure, the dynamical and thermal properties of the structure, the control of the structure, the overall spacecraft system and the cost model. (Study will be completed by end of 2017)

2. Build and operate a lab-based prototype to test the new technologies, i.e. the deployment mechanism and the control of the optics. (Complete this study by the end of 2019)
3. Design, build and launch a small-sat technology demonstrator version of SUPERSHARP. (Launch date 2022)
4. Propose SUPERSHARP for the L4 mission. (2024)
5. Start L4 mission preparation and definition phase. (2025)
6. Start L4 mission implementation phase. (2030)
7. Launch full scale SUPERSHARP L4 mission. ( 2039)

| IFU FOV | $1000$ mas $\times 1000$ mas |
|---|---|
| IFU sampling | $\sim 0.5\lambda/D$ |
| Centre of wavelength region 1 | 763nm |
| Centre of wavelength region 2 | 121nm |
| IWA at wavelength 1 | 16 mas |
| IWA at wavelength 2 | 3 mas |
| Primary mirror baseline | 24 m |
| Raw contrast at $5\times\lambda/D$ | $10^{-8}$ |
| Number of spectral channels at wavelength 1 | $\sim 50$ |
| Spectral resolution R at wavelength1 | $\sim 70$-100 |
| Number of spectral channels at wavelength 2 | $\sim 50$ |
| Spectral resolution R at wavelength 2 | $\sim 20$ |

Table 4: Summary of the top level requirements. These numbers are illustrative: a design study is needed to determine their values more accurately.

| Abbreviation | Meaning |
|---|---|
| ADI | Angular differential imaging |
| CAD | Computer aided design |
| ESA | European space agency |
| E-ELT | European extremely large telescope |
| DM | Deformable mirror |
| DMS | Dimethyl sulphide |
| FOV | field of view |
| HZ | habitable zone |
| IWA | inner working angle |
| JWST | James Webb space telescope |
| HST | Hubble space telescope |
| LUVOIR | large ultra-violet optical infra-red |
| mas | milli-arcsec |
| NIR | Near infra-red |
| PSF | Point spread function |
| RV | Radial velocity |
| SiC | Silicon carbide |
| TMT | Thirty-metre telescope |
| TRL | Technology readiness level |
| WFE | wavefront error |

Table 5: List of acronyms and abbreviations.

# 8) Lead proposer contact information

Dr. Ian R Parry
University of Cambridge
Institute of Astronomy
Madingley Rd
Cambridge
CB3 0HA
UK

+44 1223 337092

irp@ast.cam.ac.uk

www.ast.cam.ac.uk/~irp

The lead proposer can dedicate 20% of his time to support the study activities.